\documentclass[letterpaper]{article} 
\usepackage{aaai2026}  
\usepackage{times}  
\usepackage{helvet}  
\usepackage{courier}  
\usepackage[hyphens]{url}  
\usepackage{graphicx} 
\usepackage{natbib}  
\usepackage{caption} 
\usepackage{algorithm}
\usepackage{algorithmic}
\usepackage{amsmath}
\usepackage{newfloat}
\usepackage{listings}
\DeclareCaptionStyle{ruled}{labelfont=normalfont,labelsep=colon,strut=off} 
\floatstyle{ruled}
\newfloat{listing}{tb}{lst}{}
\floatname{listing}{Listing}
\title{On the Analogy between Human Brain and LLMs: Spotting Key Neurons in Grammar Perception}
\author{
    Sanaz Saki Norouzi\textsuperscript{\rm 1},
    Mohammad Masjedi\textsuperscript{\rm 2},
    Pascal Hitzler\textsuperscript{\rm 1}
}
\affiliations{
    \textsuperscript{\rm 1}Kansas State University\\\textsuperscript{\rm 2}Providence Health \& Services 

}

\usepackage{bibentry}

\begin{document}

\maketitle



\begin{abstract}

Artificial Neural Networks, the building blocks of AI, were inspired by the human brain’s network of neurons. Over the years, these networks have evolved to replicate the complex capabilities of the brain, allowing them to handle tasks such as image and language processing. In the realm of Large Language Models, there has been a keen interest in making the language learning process more akin to that of humans. While neuroscientific research has shown that different grammatical categories are processed by different neurons in the brain, we show that LLMs operate in a similar way. 

Utilizing Llama 3, we identify the most important neurons associated with the prediction of words belonging to different part-of-speech tags. Using the achieved knowledge, we train a classifier on a dataset, which shows that the activation patterns of these key neurons can reliably predict part-of-speech tags on fresh data. The results suggest the presence of a subspace in LLMs focused on capturing part-of-speech tag concepts, resembling patterns observed in lesion studies of the brain in neuroscience.

\end{abstract}

\section{Introduction}
\label{sec:intro}
As Large Language Models (LLMs) continue to show significant performance on various NLP tasks, understanding how they behave internally has gained attention in AI research. Researchers try to understand how LLMs store specific linguistic or factual knowledge, and how this knowledge can be interpreted or even controlled \cite{pezeshkpour2023measuring,zhang2025controlling}. To support these efforts, Explainable AI (XAI) techniques are employed to help researchers understand what internal components cause specific outputs and why certain predictions are made \cite{zhao2024explainability}. To this end, different explanation methods are explored such as gradient-based \cite{dai2021knowledge}, attention-based \cite{jaunet2021visqa}, and neuron-activation-based \cite{antverg2021pitfalls}. However, it is proven that attention analysis cannot be interpreted as explanations\cite{jain2019attention}. In contrast, gradient-based approaches have demonstrated greater promise in identifying important neurons \cite{chen2025identifying}. On the other hand, activation analysis methods have some limitations as raw activation values may not reliably indicate neuron importance, and there are no ground-truth explanations for the neurons based on activation \cite{zhao2024explainability}. Recent studies have primarily explored how different languages are localized inside multilingual LLMs and tried to find important neurons linked to each language or domain \cite{zhang2024unveiling, chen2025identifying}. However, the identification of specific regions related to fine-grained neuron-level grammar encoding has largely remained unexplored, especially in decoder-only models such as Llama.

In parallel, there have been studies in cognitive neuroscience to illustrate the functions of the human brain. Earlier, the researchers believed in a modular view of the brain, in which specific brain regions are associated with specific functions \cite{meunier2010modular,hagmann2008mapping}. More recent analyses provided insights into connectionism, which defines brain functions as the result of interactions between regions rather than isolated parts. In other words, a task like language understanding involves multiple regions and not just one module \cite{nazarova2015modern, arai2012effective}. However, recent studies have emphasized the integrative view which combines modularity and connectionism views. In these studies, researchers have shown that cognitive functions are not confined to a single region, but there are interactions between neural units within a specific region and across the brain \cite{perich2025neural, dragomir2021brain}.

Artificial neural networks are inspired by the human brain, and Large Language Models represent an advanced and complex form of these networks. LLMs are now utilized across a wide range of tasks and have demonstrated significant impact. Thus, understanding how they work can help us explore similarities between them and the human brain. In this paper, we provide the model with incomplete sentences and analyze its next-word predictions along with their corresponding part-of-speech (POS) tags. Using XAI approaches, we will highlight that there is an analogy between the human brain and the LLMs in terms of grammar perception. First, combining gradient-based attribution scores with statistical significance, we identify neurons that are significantly sensitive to these grammatical concepts. Next, using machine learning analyses, we show that these neurons consistently represent interpretable grammatical concepts. Overall, the key contributions of this study are:
\begin{itemize}
    \item Neural grammatical mapping in LLMs: We demonstrate neuron-level specialization for POS tags in Llama-3. 
    \item Concept-based interpretation: Using concept-based analysis and in-depth statistical methods, we provide evidence that the identified neurons encode POS tag concepts
    \item Dataset construction: In addition to the main contributions, a suitable dataset for the POS tag concept analysis is built. 
\end{itemize}

Taken together, our study shows that neurons associated with grammatical categories are distributed across all layers of the LLM, highlighting the connectionist perspective of the human brain. But more interestingly, it also indicates that there are certain neurons within each layer that are highly specialized in the perception of grammatical concepts, aligning with the latest discoveries in cognitive neuroscience about the functionality of the brain.

\section{Related Work}
\label{sec:lit}
\subsection{Knowledge Representation in the Human Brain }
Understanding how the human brain organizes and localizes linguistic knowledge has been a key question in cognitive neuroscience. Researchers have employed techniques such as brain lesion studies and neuroimaging to study how different language functions such as grammar, syntax, and meaning are handled by specific or partially distributed areas of the brain \cite{friederici2011brain}. For example, \citeauthor{shapiro2003representation}, investigated how the brain processes different grammatical categories, focusing on nouns and verbs. They discovered that the brain processes nouns and verbs differently, not only based on their meaning, but also their grammatical roles. And while some general language areas are shared, these word categories also activate distinct neural regions \cite{shapiro2003representation}. Another study introduced concept cells as selective neurons that encode abstract and invariant representations of specific concepts, 
\cite{quiroga2012concept}. In \cite{huth2016natural}, it was shown that during natural narrative listening, the brain encodes rich semantic content in spatially and semantically organized cortical maps. \citeauthor{deniz2019representation} found that there is a shared neural basis for language meaning across modalities \cite{deniz2019representation}. Another research showed that single brain neurons in humans respond 
to meaningful concepts in visual and textual format \cite{quiroga2005invariant}. Other studies have proposed that human conceptual knowledge is shaped by the integration of sense experiences and abstract language-based information, supported by distinct but interacting neural systems \cite{wang2020two, liu2025object, kiefer2012conceptual}.

\subsection{Knowledge Representation in LLMs}
LLMs have demonstrated they can handle a wide range of tasks well, which shows that they understand the structure and meaning of language \cite{minaee2024large}. Thus, researchers are trying to figure out how these models store and use knowledge. An area of investigation has focused on evaluating general factual knowledge encoded in LLMs \cite{pezeshkpour2023measuring,kadavath2022language,meng2022locating}, or domain-specific knowledge such as mathematical capabilities \cite{didolkar2024metacognitive}. In the linguistic field, \cite{liu2019linguistic, jawahar2019does}, analyzed BERT’s layer-level representations using probing tasks to investigate the extent to which they encode syntactic and structural properties of language. Other studies have investigated identifying where and how subject-verb agreement is represented in multilingual and monolingual BERT models using causal mediation analysis \cite{finlayson2021causal,mueller-etal-2022-causal}. Recent research has increasingly focused on identifying the role of individual neurons in LLMs. For example, \citeauthor{dai2021knowledge}, proposed a gradient-based method to find knowledge neurons in BERT by tracing factual outputs from cloze-style prompts \cite{dai2021knowledge}. In \cite{chen2024journey, kojima2024multilingual}, multilingual LLMs (m-BERT, m-GPT, and LLaMA 2-7B) were utilized to identify the specific neurons essential for multilingual behavior, while \cite{zhang2024unveiling} found linguistic areas that influence the handling of different languages. \cite{durrani2020analyzing} revealed that individual neurons in BERT can capture specific linguistic properties. In \cite{chen2025identifying} the authors proposed a method to calculate neuron attributions and confirm how they are responsive to the long text generated output. They did their experiments on multiple choice QA datasets to investigate how LLMs work across different domains and languages.

\section{Method}
\label{sec:method}
In this section, we outline our approach to finding an analogy between the human brain and LLMs with respect to grammar. The objective is to determine whether certain neurons in LLMs contribute more actively to generating words belonging to specific grammatical categories.

\subsection{Overview}
\begin{figure*}[t]
    \centering
    \includegraphics[width=\textwidth]{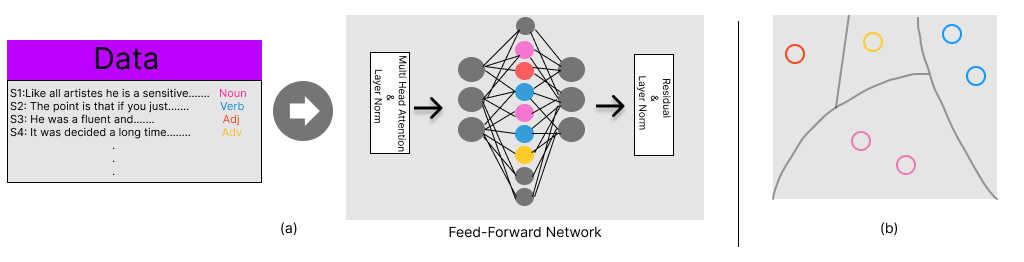}
    \caption{\centering A schematic view of the method: (a) Finding key neurons for each category -note the colors  (b) Training a classifier for concept analysis}
    \label{fig:method}
\end{figure*}

To do this, we utilize Llama-3-8B, an LLM with 32 layers and the feed-forward layer size of 14336 \cite{dubey2024llama}. We use a large dataset of partial sentences, where we split it into two parts. We start by identifying the neurons that contribute most to predicting the next word, using the first part of the dataset. We narrow our focus to four part-of-speech (POS) tag categories (verb, noun, adjective, and adverb), where we form an activation count vector for each neuron. Next, we conduct statistical tests to demonstrate that there is a significant association between some of these neurons and the POS tags. This also helps us select the most important neurons based on their p-values. Finally, by training a classifier on the second part of the dataset, we validate our assumption that certain neurons are remarkably associated with certain grammatical categories. This means that the POS tag of the next word in a partial sentence can reliably be predicted from the activation pattern of the key neurons. A high-level illustration of the proposed method is shown in Figure \ref{fig:method}, while further details are provided next.

\subsection{Step1. Finding Key Neurons}
To entertain the idea of associating certain neurons to certain grammatical concepts, we pass several partial sentences to the LLM, with the next words belonging to one of the focused categories (POS tags). To identify the key neurons responsible for each category, we first employ the Integrated Gradients method \cite{sundararajan2017axiomatic}. Integrated Gradients is an attribution method proposed to understand the predictions of deep learning models. This explainability technique aims to find which input features have contributed the most to the predicted output. As shown in Eq.\ref{eq1}, it calculates how much each input feature ($x$) affects the predicted output by slowly changing the input from baseline ($x'$, usually zero) to the actual value and adding up the changes in the prediction along the way. Being applied on the i-th feature of the input, with $\alpha$ as the interpolation factor between the baseline and the input, and F as the output function, the equation reads: 
\begin{equation}
\label{eq1}
\text{IGs}_i(x) := (x_i - x'_i) \times \int_{0}^{1} \frac{\partial F(x' + \alpha (x - x'))}{\partial x_i} \, d\alpha
\end{equation}

Besides how to select neurons, we also need to determine where in the LLM architecture this should be done. \citeauthor{geva2020transformer} found that the feed-forward networks in transformer-based language models act like key-value memory systems, helping the model predict the next word by recalling patterns from the training data \cite{geva2020transformer}. Thus, our focus in this step is on the feed-forward network. The current generation of LLMs leverages Gated Linear Units (GLUs) in their feed-forward layers \cite{dauphin2017language}. The computation is given by:
\begin{equation}
\label{eq2}
\text{Feed-Forward}(X) = (XW^{U} \odot \text{SiLU}(XW^{G})) W^{D}
\end{equation}
where the input is transformed using two different weight matrices (up-project $W^{U}$ \& gate-project $W^{G}$). The results are combined after applying the activation function to $W^{G}$, and finally passed through the down-project weight matrix ($W^{D}$). For the neuron selection, one can consider either the output of the linear function ($f=XW^{U}$) or the gated one (g = \text{SiLU}($W^{XG}$)). However, the empirical study conducted in \cite{chen2025identifying} shows that the use of neurons in the linear function provides more promising results. Thus, in our study, we also focus on this set of neurons. 

Given a partial sentence (ps) as the input to the LLM, the probability of the predicted output by the LLM is:
\begin{equation}
 P_{ps}^{\mathcal{C}}(\hat{f}^l_i) = p(\text{cat}(\hat{y}) = \mathcal{C} \mid {ps}, f^l_i = \hat{f}^l_i)   
\end{equation}
where $\mathcal{C}$ is the category of the next word (noun, verb, etc.), $\hat{y}$ is the word predicted by the model, 
${f}^l_i$ is the i-th neuron in layer l of the feedforward, and $\hat{f}^l_i$ is the value assigned to ${f}^l_i$.
Following the integrated gradients method and an its estimation based on the Riemann approximation \cite{dai2021knowledge, chen2025identifying}, the calculation of the contribution score of the neuron set is as follows, with m=10 being the number of estimation steps :
\begin{equation}
\frac{\bar{f}^l_i}{m} \sum_{k=1}^{m} \frac{\partial P_{{ps}}^{\mathcal{C}}\left( \frac{k}{m} \bar{f}^l_i \right)}{\partial f^l_i}
\end{equation}\

\textbf{Implementation:} We use the first part of our dataset to pass several partial sentences to Llama 3, and calculate the contribution of the neurons using the Integrated Gradients method. This is done for all 32 layers of the transformer, where for each sample and in each layer, the selected neurons are ranked based on their attribution score. Then we select the top n\% of the most influential neurons (with the top 5\% eventually chosen, as described in the Results section). 

The neuron selection above is based on their contribution to generating the output (which is essential); however, the selection is agnostic to part-of-speech (POS) tag categories. To address this, we categorize the selected neurons based on the POS tags of the predicted words. Then, for each layer, we select the top-k most frequent neurons within each category (noting that a neuron can appear in more than one category). This is an important step, as it ensures that a neuron is selected only if it appears in a reasonable number of samples. For instance, a neuron that exhibits a very high attribution score for a single sample but low scores across the rest is not a good candidate and will be excluded here. Additionally, selecting the top-k neurons for each POS tag diversifies the selection of neurons across different grammatical categories. We ultimately selected k= 100 as detailed in the Results section.

It should be noted that computing attribution scores for the neurons is the most computationally intensive part of the procedure. All experiments were conducted on an NVIDIA A100 GPU.

\subsection{Step2: Filtering Neurons by Statistical Significance}
The previous step provides us with a list of key neurons for each category in each layer. We can construct an activation count vector for each neuron by recording the number of times each neuron appears in the samples of each category. Having four POS tag categories in this study, the activation count vectors are of the size $1\times4$.

\begin{equation}
 V = [O_1, O_2, O_3, O_4] 
\end{equation}
where $O_i$ is the observed frequency of the $i_{th}$ category. 

Since the selection so far is based on attribution scores and the most frequently appearing neurons per category, there is still no guarantee that all these neurons have a meaningful contribution to determining the target POS tags. For instance, a neuron that consistently exhibits a high attribution score regardless of the grammatical category of the next word is not probably a suitable candidate in this context. Thus, to verify that neuron activations are not randomly distributed across syntactic concepts, we perform a Pearson's chi-squared test on the neuron's activation count vector. Two different types of test can be utilized here:

\emph{1. Goodness of fit}: This is a 1D test where the observed frequencies $O_i$ in the neuron's activation count vector are compared to the expected frequencies $E_i$ derived from the actual distribution of the POS tags in the dataset. The test statistic is
\begin{equation}
\chi^2 = \sum_{i=1}^{4} \frac{(O_i - E_i)^2}{E_i}
\end{equation}
which gets larger as the observed frequencies deviate from the expected ones, resulting in a smaller p-value. We only retain a neuron if its p-value is smaller than a threshold (e.g. 0.05), which means it is significantly associated with a certain POS tag category (or a few categories).

\emph{2. Test of independence}: This is a 2D test where the number of times a neuron is active for each category, and the number of times it is inactive for that category are put together in a $2\times4$ contingency table. The test statistic is
\begin{equation}
\chi^2 = \sum_{i=1}^{4} \sum_{j=1}^{2} \frac{(O_{ij} - E_{ij})^2}{E_{ij}}
\end{equation}
with p-values below a certain threshold highlighting a significant dependency between the neuron's identity and predicted POS tags. 

Regardless of which method is chosen, performing the statistical significance test at this stage is crucial to filtering the neurons and retaining only those with a strong association with one (or a few) grammatical categories.

\textbf{Implementation:} We directly apply chi-squared test to the activation count vectors and select a subset of neurons by their statistical significance. We ultimately chose the first method, using a p-value threshold of 0.05, as described in the Results section.

\subsection{Step3: Concept Activation Analysis}
To investigate whether interpretable grammatical concepts are encoded within the internal representations of LLMs, we employ concept-based techniques commonly used in XAI to analyze model behavior in terms of high-level, human-interpretable abstractions.
In particular, we leverage Concept Activation Vectors (CAV) \cite{kim2018interpretability} and Concept Activation Regions (CAR) \cite{crabbe2022concept}, the techniques that test whether a concept is detectably represented in the hidden activation space of a model. These methods enable us to go beyond surface-level attributions and instead interrogate whether specific groups of neurons consistently encode grammatical concepts such as POS tags. 

To accomplish this, we leverage insights gained previously (identifying key neurons associated with the grammatical concepts), to train classifiers on new, unseen data. The goal is to explore the possibility of predicting the POS tag of the next word, solely by knowing the activation patterns of the key neurons. In this context, a linear decision boundary represents the Concept Activation Vector (CAV), while a non-linear boundary represents the Concept Activation Region (CAR).

\textbf{Implementation:} After finalizing the selection of the most influential neurons in the previous step, we use the second part of the dataset (which was not used before), to perform the concept activation analysis. Consistent with Step 1, we apply Integrated Gradients method to this part of the dataset to spot the top contributing neurons based on their attribution scores. To show whether or not these neurons are among the final selection of neurons identified previously, we apply one-hot encoding. The length of the feature vector is equal to the number of key neurons identified in the previous step, with its elements indicating the presence (1) or absence (0) of those neurons among the top neuron in the current step. The target variable is the POS tag of the next word for each sample. 

We split the data into train and test sets (80\% and 20\%, respectively), and train a Support Vector Machine (SVM) classifier on the training portion. We choose SVM because it supports both linear and nonlinear decision boundaries through using linear and nonlinear kernels (to highlight both Concept Activation Vector (CAV) and Concept Activation Region (CAR)). Machine learning analyses are conducted independently across all 32 layers, with both linear and nonlinear kernel setup. As for the evaluation metrics, we compute accuracy, precision, recall, and F1 score, However, we ultimately choose accuracy along with macro-averaged F1 score to compare different configurations of the model. A reasonable evaluation score (particularly on the test set) can indicate the potential to predict the grammatical category of the next word in a partial sentence, solely based on the activation behavior of the key neurons. Further details regarding the setup and the findings will be provided in the Results section.

Additionally, to further demonstrate that the identified neurons can genuinely  capture part-of-speech (POS) tag concepts rather than reflecting arbitrary patterns, we conduct an ablation study. We compare the effect of disabling the key neurons against disabling a random selection of neurons. The objective is to assess how often the predicted next words differ in grammatical category (POS tag) from those in the original predictions. For more details, see the Results section.

\section{Experiments and Analysis}
\label{sec:res}

\subsection{Dataset Construction}
We use the British National Corpus (BNC\footnote{http://www.natcorp.ox.ac.uk/}) \cite{20.500.14106/2554} to generate a dataset for evaluating our approach. BNC is a 100 million word dataset that includes about six million sentences with their corresponding part-of-speech (POS) tags. To create our dataset, we first split the sentences into different lengths, making sure each split stays within the valid index range (with the maximum index being \mbox{length -- 1}). This process gives us a dataset of 20 million partial sentences. For our experiments, we filter the partial sentences to include only those with a minimum length of three, then randomly sample 500,000 from that set.

Next, we use an LLM (GPT-4o-mini) to generate 10 diverse and meaningful completions for each partial sentence. The reason for using LLM-generated completions here is to ensure that the final dataset for our study consists solely of partial sentences that are followed by a unique and consistent POS tag for the next word. For example, if the actual next word in the database is a noun, but the LLM has predicted any next word other than noun in any of its ten completions, we exclude that partial sentence from our final selection. To do this, we use SpaCy \cite{honnibal2020spacy} to extract the POS tags of the next word in the LLM-generated completions, and list them along with the POS tag of the actual next word in the BNC. To be consistent on the labels, we standardize the labels by mapping them to their super categories (for example, grouping 'NN0', 'NN1', 'NN2', etc., under the general label 'Noun'). We then restrict our dataset to partial sentences where the POS tag for the next word remains consistent between the original data and all the LLM predictions. Although this significantly reduces the number of viable candidates, it is an essential step to ensure that the input to our study is appropriate. Also, since our focus is on four primary POS tags (noun, verb, adjective, and adverb), we refine our final selection to sentences that fall within these categories.

For our experiments, we finally select a subset of data with a sample size roughly proportional to the number of available examples in each class: 2000 for Nouns, 2000 for Verbs, 500 for Adjectives, and 300 for Adverbs. Half of these samples are used to spot the key neurons (Step 1 and 2 of the Method section), while the the other half are used to train and test classifiers using an 80\%-20\% split (Step 3 of the Method section).

It is essential to emphasize that none of the data we eventually select for our experiments includes the LLM-generated completions or words mentioned above. The data is sourced exclusively from the BNC dataset, with the LLM serving only to assist in selecting concrete samples for our analysis.

\subsection{Results}
As mentioned earlier, we apply the Integrated Gradients method to the first part of our final dataset to identify and retain neurons with the highest attribution scores. At this stage, for each sample, we retain the top 5\% of the neurons in every 32 layers of Llama 3 (about 700 out of 14336 neurons). It should be noted that we also tried selecting more neuron (e.g. top 10\%), but it revealed to be unnecessary. This is because we do further refinement in the following steps, and selecting more neurons here will not substantially change the outcome.

Next, for each class (POS tag), we select the top-k most frequently occurring neurons across all samples. To determine the suitable value for k, we experimented with multiple configurations (k = 50, 100, 200). Taking into account the statistical significance test performed in the following step, we concluded that choosing k = 100 here provided a good balance between including sufficiently informative features and avoiding redundancy. The exact number of neurons for each layer after this step is shown in Table \ref{tab:svm-results} (Column \# N). Note that the results for other k values are provided in the supplementary material.

We then construct a $1\times4$ count vector for each key neuron 
(in each layer), recording the number of times it appears in the samples of each class. Next, we conduct a chi-squared test on each vector to assess the discriminative power of the selected neurons, and retain only those with statistical significance. As mentioned in the Method section, we conduct two different types of chi-squared test. A goodness of fit test which is a 1D test ($1\times4$) with DOF = 3, and test of independence which is a 2D test ($2\times4$) with DOF = 3. Although the two methods are fundamentally different, the results reveal that both of them rank the neuron in a similar (but not identical) order. However, at the same significance level ($\alpha$ = 0.05), the goodness of fit test narrows the selection to a smaller subset while maintaining accuracy of the subsequent SVM. Therefore, we ultimately choose that test for neuron selection. Note that the final results using the other test (test of independence) are provided in the supplementary material.
The number of selected neurons for each layer after conducting the test is shown in Table \ref{tab:svm-results} (Column \# Sel N).

\begin{figure}[t]
    \centering
    \includegraphics[width=\linewidth]{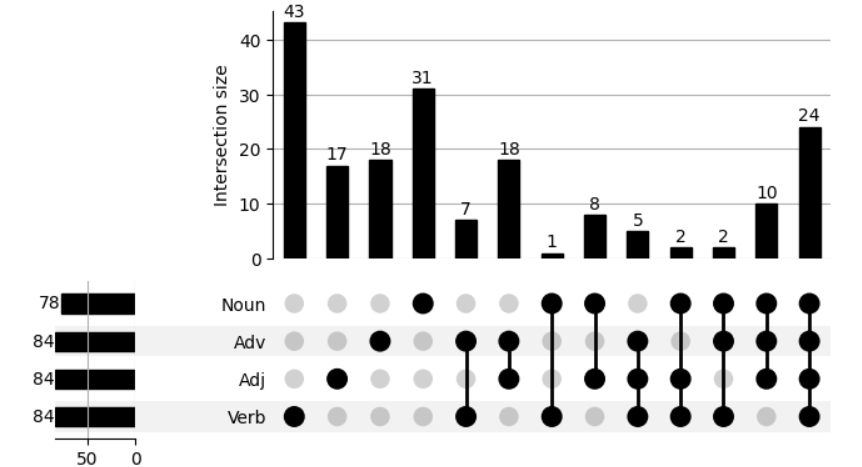}
    \caption{\centering Overlap between selected neurons (p-value $<$ 0.05) for Layer 20 of Llama-3}
    \label{fig:neuron}
\end{figure}

Figure \ref{fig:neuron} shows the number of selected neurons for a random layer (Layer 20). It illustrates that each POS tag has activated some unique neurons, while some other neurons are shared across multiple tags. This pattern aligns with the findings in neuroscience about specialization and overlap within the brain.

Thereafter, we train SVM classifiers on the second part of our final dataset (that is not used previously). The classifiers are trained separately for each of the 32 layers' key neurons. As stated earlier, the desired outcome is to be able to predict the target variable (POS tag category of the next word), solely by tracing the activation pattern of the key neurons. The SVM is trained using both linear and nonlinear kernels to represent the CAV and the CAR methods. It is also carried out using the selected neurons, both before and after the chi-squared test.

The SVM results are summarized in Table \ref{tab:svm-results}, where the accuracy score, along with the macro-averaged F1 score are shown for each layer. It is observed that CAV and CAR exhibit similar performance, with no significant difference between them in terms of the evaluation metrics. Also, it is evident that the neuron refinement using the chi-squared test has preserved the evaluation scores, thereby supporting the effectiveness of the test in eliminating neurons less associated with grammatical concepts, thus constructing a smaller and more efficient model with fewer parameters. 

\begin{table*}[ht]
\centering
\small  
\renewcommand{\arraystretch}{1.2}
\resizebox{\textwidth}{!}{
\begin{tabular}{|c|c|c|c|c|c|c|c|c|c|c|}
\hline
Layer & CAV ACC & CAV F1 & CAR ACC & CAR F1 & CAV-Sel ACC & CAV-Sel F1 & CAR-Sel ACC & CAR-Sel F1 & \# N & \# Sel N \\
\hline
Layer 1  & 0.77 (0.82) & 0.71 (0.77) & 0.73 (0.74) & 0.65 (0.71) & 0.77 (0.82) & 0.71 (0.77) & 0.73 (0.74) & 0.65 (0.71) & 223 & 181 \\
Layer 2  & 0.74 (0.77) & 0.68 (0.75) & 0.75 (0.85) & 0.68 (0.86) & 0.74 (0.77) & 0.68 (0.75) & 0.75 (0.85) & 0.68 (0.86) & 201 & 160 \\
Layer 3  & 0.70 (0.78) & 0.64 (0.75) & 0.69 (0.73) & 0.61 (0.73) & 0.70 (0.78) & 0.64 (0.75) & 0.69 (0.73) & 0.61 (0.73) & 202 & 152 \\
Layer 4  & 0.79 (0.81) & 0.73 (0.78) & 0.75 (0.77) & 0.71 (0.75) & 0.79 (0.81) & 0.73 (0.78) & 0.75 (0.77) & 0.71 (0.75) & 198 & 157 \\
Layer 5  & 0.73 (0.82) & 0.65 (0.80) & 0.72 (0.77) & 0.68 (0.75) & 0.79 (0.81) & 0.73 (0.78) & 0.72 (0.77) & 0.68 (0.75) & 217 & 176 \\
Layer 6  & 0.75 (0.83) & 0.68 (0.80) & 0.73 (0.78) & 0.67 (0.74) & 0.75 (0.83) & 0.68 (0.80) & 0.73 (0.78) & 0.67 (0.74) & 211 & 163 \\
Layer 7  & 0.76 (0.82) & 0.69 (0.80) & 0.75 (0.78) & 0.70 (0.77) & 0.76 (0.82) & 0.69 (0.80) & 0.75 (0.78) & 0.70 (0.77) & 226 & 165 \\
Layer 8  & 0.71 (0.79) & 0.61 (0.75) & 0.71 (0.76) & 0.64 (0.73) & 0.71 (0.79) & 0.61 (0.75) & 0.71 (0.76) & 0.64 (0.73) & 209 & 146 \\
Layer 9  & 0.71 (0.79) & 0.60 (0.72) & 0.72 (0.77) & 0.60 (0.70) & 0.71 (0.79) & 0.60 (0.72) & 0.72 (0.77) & 0.60 (0.70) & 174 & 89 \\
Layer 10 & 0.69 (0.76) & 0.61 (0.71) & 0.68 (0.73) & 0.58 (0.68) & 0.69 (0.76) & 0.61 (0.71) & 0.68 (0.73) & 0.58 (0.68) & 183 & 101 \\
Layer 11 & 0.70 (0.78) & 0.60 (0.74) & 0.72 (0.77) & 0.64 (0.72) & 0.70 (0.78) & 0.60 (0.74) & 0.72 (0.77) & 0.64 (0.72) & 183 & 99 \\
Layer 12 & 0.78 (0.81) & 0.70 (0.75) & 0.76 (0.81) & 0.68 (0.77) & 0.78 (0.81) & 0.70 (0.75) & 0.76 (0.81) & 0.68 (0.77) & 196 & 105 \\
Layer 13 & 0.81 (0.87) & 0.70 (0.80) & 0.81 (0.84) & 0.71 (0.79) & 0.81 (0.87) &  0.70 (0.80) & 0.81 (0.84) & 0.71 (0.79) & 180 & 120 \\
Layer 14 & 0.84 (0.88) & 0.73 (0.83) & 0.85 (0.85) & 0.75 (0.80) & 0.84 (0.88) & 0.73 (0.83) & 0.85 (0.85) & 0.75 (0.80) & 196 & 153 \\
Layer 15 & 0.82 (0.91) & 0.71 (0.86) & 0.86 (0.89) & 0.79 (0.83) & 0.82 (0.91) & 0.71 (0.86) & 0.86 (0.89) & 0.79 (0.83) & 202 & 165 \\
Layer 16 & 0.89 (0.92) & 0.80 (0.87) & 0.89 (0.91) & 0.80 (0.86) & 0.89 (0.92) & 0.80 (0.87) & 0.89 (0.91) & 0.80 (0.86) & 203 & 175 \\
Layer 17 & 0.88 (0.91) & 0.77 (0.86) & 0.87 (0.90) & 0.77 (0.84) & 0.88 (0.91) & 0.77 (0.86) & 0.87 (0.90) & 0.77 (0.84) & 208 & 184 \\
Layer 18 & 0.90 (0.93) & 0.80 (0.89) & 0.89 (0.92) & 0.80 (0.88) & 0.90 (0.93) & 0.80 (0.89) & 0.89 (0.92) & 0.80 (0.88) & 198 & 181 \\
Layer 19 & 0.89 (0.94) & 0.77 (0.90) & 0.90 (0.92) & 0.78 (0.88)  & 0.89 (0.94) & 0.77 (0.90) & 0.90 (0.92) & 0.78 (0.88) & 211 & 192 \\
Layer 20 & 0.91 (0.94) & 0.83 (0.89) & 0.90 (0.92) & 0.81 (0.87) & 0.91 (0.94) & 0.83 (0.89) & 0.90 (0.92) & 0.81 (0.87) & 209 & 186 \\
Layer 21 & 0.93 (0.95) & 0.87 (0.92) & 0.92 (0.93) & 0.86 (0.90) & 0.93 (0.95) & 0.87 (0.92) & 0.92 (0.93) & 0.86 (0.90) & 242 & 220 \\
Layer 22 & 0.94 (0.96) & 0.87 (0.93) & 0.93 (0.95) & 0.84 (0.92) & 0.94 (0.96) & 0.87 (0.93) & 0.93 (0.95) & 0.84 (0.92) & 223 & 206 \\
Layer 23 & 0.94 (0.95) & 0.89 (0.92) & 0.92 (0.94) & 0.86 (0.90) & 0.94 (0.95) & 0.89 (0.92) & 0.92 (0.94) & 0.86 (0.90) & 235 & 222 \\
Layer 24 & 0.93 (0.95) & 0.88 (0.91) & 0.94 (0.94) & 0.88 (0.91) & 0.93 (0.95) & 0.88 (0.91) & 0.94 (0.94) & 0.88 (0.91) & 246 & 231 \\
Layer 25 & 0.93 (0.95) & 0.85 (0.92) & 0.92 (0.94) & 0.84 (0.91) & 0.93 (0.95) & 0.85 (0.92) & 0.92 (0.94) & 0.84 (0.91) & 244 & 229 \\
Layer 26 & 0.92 (0.96) & 0.83 (0.92) & 0.93 (0.94) & 0.86 (0.90) & 0.92 (0.96) & 0.83 (0.92) & 0.93 (0.94) & 0.86 (0.90) & 219 & 198 \\
Layer 27 & 0.91 (0.95) & 0.81 (0.91) & 0.91 (0.94) & 0.83 (0.89) & 0.91 (0.95) & 0.81 (0.91) & 0.91 (0.94) & 0.83 (0.89) & 218 & 192 \\
Layer 28 & 0.94 (0.96) & 0.88 (0.93) & 0.95 (0.94) & 0.90 (0.91) & 0.94 (0.96) & 0.88 (0.93) & 0.95 (0.94) & 0.90 (0.91) & 232 & 200 \\
Layer 29 & 0.93 (0.94) & 0.87 (0.90) & 0.93 (0.93) & 0.85 (0.89) & 0.93 (0.94) & 0.87 (0.90) & 0.93 (0.93) & 0.85 (0.89) & 192 & 146 \\
Layer 30 & 0.91 (0.94) & 0.80 (0.90) & 0.91 (0.94) & 0.78 (0.90) & 0.91 (0.94) & 0.80 (0.90) & 0.91 (0.94) & 0.78 (0.90) & 190 & 153 \\
Layer 31 & 0.91 (0.92) & 0.83 (0.86) & 0.91 (0.91) & 0.82 (0.85) & 0.91 (0.92)  & 0.83 (0.86) & 0.91 (0.91) & 0.82 (0.85) & 150 & 84 \\
Layer 32 & 0.90 (0.94) & 0.82 (0.91) & 0.91 (0.92) & 0.84 (0.89) & 0.90 (0.94) & 0.82 (0.91) & 0.91 (0.92) & 0.84 (0.89) & 202 & 97 \\
\hline
\end{tabular}
}
\caption{Accuracy and macro-avg F1 scores of the test set (training set) across all layer of Llama-3 for CAV and CAR methods, before and after neuron selection by the chi-squared test.}
\label{tab:svm-results}
\end{table*}
Overall, the results from the SVM are promising, and supports our assumption that certain neurons have strong association with certain grammatical categories. To dive deeper, we focus on the CAV method after neuron selection by the chi-squared test (CAV-Sel in Table \ref{tab:svm-results}). The mean accuracy of the test set across all 32 layers is 0.84, and the mean macro-avg F1 score is 0.76. On the last layer (Layer 32), the accuracy and F1 score are 0.91 and 0.84, respectively. This means that by focusing on the activation patterns of just 97 neurons in the last layer (out of 14336 neurons in that layer), and completely disregarding the other layers, one can correctly predict the POS tag of the next word in 91 out of 100 unseen samples.

Interestingly, as we move deeper into the layers, the scores improve gradually (though not strictly monotonic). For example, in CAV-Sel in Table \ref{tab:svm-results}, the mean accuracy score of the test set for the first 5 layers is 0.73, while it is 0.92 for the last 5 layers. This suggests that the association between the activation behavior of neurons and the grammatical concepts became increasingly distinct and well-defined as we move deeper into the transformer's layers.

To further validate that the identified neurons are truly related to the grammatical concepts, we also conduct an ablation study on the developed model, and compare the results to those of a control experiment. To do this, we collect 200 test samples for each grammatical category. We then disable the identified key neurons in the model and measure the drop in the POS tag prediction performance (which we call distortion). This means how often the predicted next word changes its grammatical category (POS tag) compared to the original prediction before disabling the neurons. As a control experiment, we repeat this by disabling an equal number of randomly selected neurons in each layer. The result distortions in terms of percentage are shown in Table \ref{tab:PD}. These results are significant, especially considering that there are only 5228 identified key neurons across all 32 layers (out of $14336\times$32 = 458752 neurons). This means that disabling our identified key neurons (which account for only 1.1\% of the total neurons) results in an average 41.2\% change in the grammatical category of the next predicted word. On the other hand, the average drop for the control experiment is only about 3.6\%. 

Collectively, our results provide strong evidence that some neurons (which are scattered across different layers) are functionally specialized and play a crucial role in encoding grammatical information. These findings are aligned with the hypothesis that LLMs, like the human brain, exhibit both localized and distributed representations for abstract linguistic functions such as grammar (i.e. POS tags).

\begin{table}[t]
\centering
\footnotesize 
\setlength{\tabcolsep}{4pt} 
\renewcommand{\arraystretch}{1.2} 
\begin{tabular}{|c|c|c|}
\hline
\textbf{Tag} & \textbf{Distortion-key neurons} & \textbf{Distortion-random neurons} \\
\hline
Verb & 46.3\% & 4.5\% \\
\hline
Noun & 22.3\% & 4.9\% \\
\hline
Adjective & 59.2\% & 4.1\% \\
\hline
Adverb & 37\% & 0.85\% \\

\hline
\end{tabular}
\caption{POS tag change when ablating key neurons vs random neurons }
\label{tab:PD}
\end{table}

\section{Conclusion and Future Work}
In this paper, we aimed to find an analogy the between human brain and LLMs in terms of grammar perception. This contributes to explaining their behavior through human understandable interpretations, and is a step towards making their generative process transparent. To this end, we constructed a dataset of incomplete sentences, each labeled with the POS tag of its predicted next word. We applied the Integrated Gradients method along with statistical tests to spot the key neurons associated with grammar understanding in Llama-3. Then using concept-based interpretations, we proposed that there is a subspace in LLMs, consisting of certain neurons scattered across all layers, that is functionally specialized in understanding grammar. Our findings are aligned with the recent discoveries in cognitive neuroscience about localized and shared functional regions of the brain.

As for the future line of research, we plan to study how LLMs represent semantics, focusing on abstract and concrete concepts, and investigate if they organize these concepts in a structured way like human brain. Moreover, the knowledge gained from our experiment and similar studies can help fine-tune LLMs for specific tasks more efficiently, and control their output to operate more naturally and in closer alignment with human understanding.

\newpage

\bibliography{aaai2026}


\section{Appendix}

The SVM results for the additional parameters discussed in the paper are presented below.
\begin{table*}[ht]
\small  
\renewcommand{\arraystretch}{1.2}
\resizebox{\textwidth}{!}{
\begin{tabular}{|c|c|c|c|c|c|c|c|c|c|c|}
\hline
Layer & CAV ACC & CAV F1 & CAR ACC & CAR F1 & CAV-Sel ACC & CAV-Sel F1 & CAR-Sel ACC & CAR-Sel F1 & \# N & \# Sel N \\
\hline
Layer 1  & 0.80 (0.87) & 0.71 (0.84) & 0.74 (0.78) & 0.65 (0.77)& 0.80 (0.87) & 0.71 (0.84) & 0.74 (0.78) & 0.65 (0.77) & 416 & 339 \\
Layer 2  & 0.76 (0.86) & 0.69 (0.84) & 0.74 (0.77) & 0.68 (0.78) &  0.76 (0.86)& 0.69 (0.84) & 0.74 (0.77) & 0.68 (0.78) & 425 & 341\\
Layer 3  & 0.74 (0.86) & 0.65 (0.85) & 0.73 (0.79) & 0.65 (0.80) & 0.74 (0.86) & 0.65 (0.85) & 0.73 (0.79)& 0.65 (0.80) & 395 & 300 \\
Layer 4  & 0.76 (0.87) & 0.68 (0.86)& 0.75 (0.78) & 0.72 (0.79) & 0.76 (0.87) & 0.68 (0.86) & 0.75 (0.78) &  0.72 (0.79) & 403 & 311 \\
Layer 5  & 0.78 (0.90) & 0.72 (0.88) & 0.72 (0.80) & 0.67 (0.79)&  0.78 (0.90) & 0.72 (0.88) & 0.72 (0.80) & 0.67 (0.79) & 420 & 328 \\
Layer 6  & 0.78 (0.89) & 0.68 (0.87) & 0.73 (0.79) & 0.66 (0.80)& 0.78 (0.89) &  0.68 (0.87) & 0.73 (0.79) & 0.66 (0.80) & 404 & 318 \\
Layer 7  & 0.78 (0.89)& 0.70(0.87)& 0.74 (0.81) & 0.68 (0.80) & 0.78 (0.89) & 0.70(0.87)& 0.74 (0.81) & 0.68 (0.80) & 424 & 299 \\
Layer 8  & 0.71 (0.86) & 0.61 (0.85) & 0.70 (0.79) & 0.60 (0.77)&  0.71 (0.86) & 0.61 (0.85) & 0.70 (0.79) & 0.60 (0.77) & 402 & 267 \\
Layer 9  & 0.74 (0.87)& 0.63 (0.84) & 0.74 (0.83) & 0.62 (0.80) & 0.74 (0.87)& 0.63 (0.84) & 0.74 (0.83) & 0.62 (0.80)& 356 & 183 \\
Layer 10 & 0.72 (0.86) & 0.63 (0.83)& 0.68 (0.78) & 0.61 (0.76)& 0.72 (0.86) & 0.63 (0.83)& 0.68 (0.78) & 0.61 (0.76) & 360 & 195 \\
Layer 11 & 0.74 (0.86) & 0.63 (0.84) & 0.71 (0.82) & 0.61 (0.78) & 0.74 (0.86) & 0.63 (0.84) & 0.71 (0.82) & 0.61 (0.78) & 355 & 186 \\
Layer 12 & 0.79 (0.88) & 0.70 (0.85) & 0.79 (0.85) & 0.73 (0.83)& 0.79 (0.88) & 0.70 (0.85) & 0.79 (0.85) & 0.73 (0.83) & 381 & 197 \\
Layer 13 & 0.83 (0.91) & 0.73 (0.87)& 0.82 (0.87) & 0.73 (0.84)& 0.83 (0.91) & 0.73 (0.87)& 0.82 (0.87) & 0.73 (0.84) & 368 & 225 \\
Layer 14 & 0.86 (0.93)& 0.75 (0.91& 0.86 (0.90) & 0.76 (0.86) & 0.86 (0.93)& 0.75 (0.91& 0.86 (0.90) & 0.76 (0.86) & 399 & 311 \\
Layer 15 & 0.86 (0.94) & 0.77 (0.93) & 0.88 (0.91) & 0.80 (0.88)& 0.86 (0.94) & 0.77 (0.93) & 0.88 (0.91) & 0.80 (0.88) & 413 & 326 \\
Layer 16 & 0.90 (0.95) & 0.83 (0.92)& 0.91 (0.93)& 0.83 (0.89)&  0.90 (0.95) & 0.83 (0.92)& 0.91 (0.93)& 0.83 (0.89)& 422 & 353 \\
Layer 17 & 0.90 (0.96) & 0.81 (0.93) & 0.89 (0.92) & 0.81 (0.89) & 0.90 (0.96) & 0.81 (0.93) & 0.89 (0.92) & 0.81 (0.89) & 428 & 381 \\
Layer 18 & 0.91 (0.95)& 0.83 (0.93) & 0.91 (0.93) & 0.86 (0.90) & 0.91 (0.95)& 0.83 (0.93) & 0.91 (0.93) & 0.86 (0.90)& 414 & 371\\
Layer 19 & 0.91 (0.97) & 0.80 (0.96)& 0.90 (0.94)& 0.79 (0.91) & 0.91 (0.97) & 0.80 (0.96)& 0.90 (0.94)& 0.79 (0.91)& 453 & 422 \\
Layer 20 & 0.93 (0.97) & 0.85 (0.96) & 0.90 (0.94) & 0.82 (0.92) & 0.93 (0.97) & 0.85 (0.96) & 0.90 (0.94) & 0.82 (0.92) & 462 & 439 \\
Layer 21 & 0.93 (0.97) & 0.88 (0.95)& 0.94 (0.95) & 0.90 (0.93) & 0.93 (0.97) & 0.88 (0.95)& 0.94 (0.95) & 0.90 (0.93) & 493 & 451 \\
Layer 22 & 0.95 (0.98) & 0.89 (0.96) & 0.93 (0.96) & 0.86 (0.95)& 0.95 (0.98) & 0.89 (0.96) & 0.93 (0.96) & 0.86 (0.95) & 463 & 436 \\
Layer 23 & 0.95 (0.97) & 0.91 (0.96)& 0.93 (0.95)& 0.88 (0.93) &  0.95 (0.97) & 0.91 (0.96)& 0.93 (0.95)& 0.88 (0.93) & 501 & 477 \\
Layer 24 & 0.94 (0.97) & 0.89 (0.96)& 0.95 (0.96) & 0.91 (0.94)& 0.94 (0.97) & 0.89 (0.96)& 0.95 (0.96) & 0.91 (0.94) & 502 & 478 \\
Layer 25 & 0.94 (0.97)& 0.89 (0.96)& 0.93 (0.95) & 0.88 (0.93) & 0.94 (0.97)& 0.89 (0.96)& 0.93 (0.95) & 0.88 (0.93) & 496 & 464 \\
Layer 26 & 0.93 (0.98) & 0.87 (0.96) & 0.92 (0.95) & 0.85 (0.93) & 0.93 (0.98) & 0.87 (0.96) & 0.92 (0.95) & 0.85 (0.93) & 462 & 427 \\
Layer 27 & 0.93 (0.98) & 0.86 (0.97)& 0.93 (0.96) & 0.87 (0.94)& 0.93 (0.98) & 0.86 (0.97)& 0.93 (0.96) & 0.87 (0.94) & 433 & 386 \\
Layer 28 & 0.95 (0.98) & 0.90 (0.97)& 0.95 (0.96) & 0.91 (0.95) & 0.95 (0.98) & 0.90 (0.97)& 0.95 (0.96) & 0.91 (0.95)  & 444& 384 \\
Layer 29 & 0.95 (0.97) & 0.91 (0.96)& 0.92 (0.94) & 0.87 (0.93)& 0.95 (0.97) & 0.91 (0.96)& 0.92 (0.94) & 0.87 (0.93) & 388 & 323 \\
Layer 30 & 0.94 (0.97) & 0.85 (0.95) & 0.93 (0.95)& 0.85 (0.93) & 0.94 (0.97) & 0.85 (0.95) & 0.93 (0.95)& 0.85 (0.93) & 387 & 320 \\
Layer 31 & 0.94 (0.96) & 0.88 (0.93)& 0.94 (0.95)& 0.88 (0.93)&  0.94 (0.96) & 0.88 (0.93)& 0.94 (0.95)& 0.88 (0.93) & 312 & 225 \\
Layer 32 & 0.93 (0.96) & 0.87 (0.95) & 0.92 (0.94) & 0.83 (0.92)& 0.93 (0.96) & 0.87 (0.95) & 0.92 (0.94) & 0.83 (0.92) & 286& 149 \\

\hline
\end{tabular}
}
\caption{Accuracy and macro-avg F1 scores of the test set (training set) for CAV and CAR methods, before and after neuron selection by the chi-squared test for K=200}
\label{tab:large_table}
\end{table*}
\vspace*{\fill}

\begin{table*}[hb]
\small  
\renewcommand{\arraystretch}{1.2}
\resizebox{\textwidth}{!}{
\begin{tabular}{|c|c|c|c|c|c|c|c|c|c|c|}
\hline
Layer & CAV ACC & CAV F1 & CAR ACC & CAR F1 & CAV-Sel ACC & CAV-Sel F1 & CAR-Sel ACC & CAR-Sel F1 & \# N & \# Sel N \\
\hline
Layer 1  & 0.71 (0.73) & 0.62 (0.67) & 0.68 (0.68)& 0.60 (0.62)&0.71 (0.73) & 0.62 (0.67) & 0.68 (0.68)& 0.60 (0.62) & 119 & 92 \\
Layer 2  & 0.66 (0.68) &  0.58 (0.64)& 0.67 (0.67) & 0.61 (0.65)& 0.66 (0.68) &  0.58 (0.64)& 0.67 (0.67) & 0.61 (0.65)& 106 & 81\\
Layer 3  & 0.62 (0.65) & 0.56 (0.61) & 0.63 (0.66) & 0.58 (0.63)& 0.62 (0.65) & 0.56 (0.61) & 0.63 (0.66) & 0.58 (0.63)& 91 & 60 \\
Layer 4  & 0.75 (0.75) & 0.65 (0.70) & 0.73 (0.75)& 0.65 (0.71)&0.75 (0.75) & 0.65 (0.70) & 0.73 (0.75)& 0.65 (0.71)& 93 & 74 \\
Layer 5  & 0.72 (0.77) & 0.63 (0.70) & 0.71 (0.75) & 0.66 (0.70)&0.72 (0.77) & 0.63 (0.70) & 0.71 (0.75) & 0.66 (0.70) & 111  & 91\\
Layer 6  & 0.70 (0.75) & 0.60 (0.71) & 0.69 (0.73) & 0.62 (0.70)& 0.70 (0.75) & 0.60 (0.71) & 0.69 (0.73) & 0.62 (0.70)& 107 & 81 \\
Layer 7  & 0.74 (0.78) &0.69 (0.73)& 0.73(0.76)& 0.67 (0.71) & 0.74 (0.78) &0.69 (0.73)& 0.73(0.76)& 0.67 (0.71) & 108 & 80 \\
Layer 8  & 0.66 (0.72) & 0.58 (0.67) & 0.66 (0.72) & 0.59 (0.67)&0.66 (0.72) & 0.58 (0.67) & 0.66 (0.72) & 0.59 (0.67)& 107 & 77 \\
Layer 9  & 0.68 (0.72) &  0.58 (0.64) & 0.68 (0.73) & 0.54 (0.65)&0.68 (0.72) &  0.58 (0.64) & 0.68 (0.73) & 0.54 (0.65) & 89 & 46 \\
Layer 10 & 0.69 (0.70) & 0.60 (0.64) & 0.67 (0.69) & 0.56 (0.62)&0.69 (0.70) & 0.60 (0.64) & 0.67 (0.69) & 0.56 (0.62)& 104 & 58 \\
Layer 11 & 0.65 (0.72) & 0.55 (0.65) & 0.66 (0.72) & 0.54 (0.65) &0.65 (0.72) & 0.55 (0.65) & 0.66 (0.72) & 0.54 (0.65)& 87 & 48 \\
Layer 12 & 0.71 (0.75)& 0.59 (0.67)& 0.68 (0.74) & 0.56 (0.67) &0.71 (0.75)& 0.59 (0.67)& 0.68 (0.74) & 0.56 (0.67) & 99 & 54 \\
Layer 13 & 0.75 (0.78) & 0.63 (0.67) & 0.75 (0.79) & 0.64 (0.70) & 0.75 (0.78) & 0.63 (0.67) & 0.75 (0.79) & 0.64 (0.70) & 88 & 58 \\
Layer 14 & 0.77 (0.80) & 0.65 (0.73) & 0.80 (0.79) & 0.71 (0.74)& 0.77 (0.80) & 0.65 (0.73) & 0.80 (0.79) & 0.71 (0.74)& 100 & 79 \\
Layer 15 & 0.80 (0.84) & 0.67 (0.78) & 0.81 (0.84) & 0.71 (0.78) &0.80 (0.84) & 0.67 (0.78) & 0.81 (0.84) & 0.71 (0.78)& 102 &  82\\
Layer 16 & 0.86 (0.88) & 0.77 (0.80) & 0.86 (0.88) & 0.74 (0.81)&0.86 (0.88) & 0.77 (0.80) & 0.86 (0.88) & 0.74 (0.81)& 105  &  91\\
Layer 17 & 0.84 (0.87)& 0.72 (0.79)& 0.83 (0.86)& 0.70 (0.79) &0.84 (0.87)& 0.72 (0.79)& 0.83 (0.86)& 0.70 (0.79)& 100 &  87\\
Layer 18 & 0.86 (0.88) & 0.75 (0.81)& 0.87 (0.88) & 0.76 (0.80) &0.86 (0.88) & 0.75 (0.81)& 0.87 (0.88) & 0.76 (0.80)& 97 & 86\\
Layer 19 & 0.86 (0.90) & 0.74 (0.83) & 0.88 (0.89) &  0.76 (0.83)&0.86 (0.90) & 0.74 (0.83) & 0.88 (0.89) &  0.76 (0.83)& 103 &  89\\
Layer 20 & 0.88 (0.91) & 0.79 (0.85) & 0.89 (0.90)& 0.80 (0.83)&0.88 (0.91) & 0.79 (0.85) & 0.89 (0.90)& 0.80 (0.83)& 107 &  93\\
Layer 21 & 0.90 (0.92) & 0.83 (0.87) & 0.90 (0.95) & 0.83 (0.94) &0.90 (0.92) & 0.83 (0.87) & 0.90 (0.95) & 0.83 (0.94)& 111 & 97 \\
Layer 22 & 0.92 (0.93)& 0.83 (0.88) & 0.91 (0.92) & 0.83 (0.86)& 0.92 (0.93)& 0.83 (0.88) & 0.91 (0.92) & 0.83 (0.86)& 117 & 106 \\
Layer 23 & 0.92 (0.93) & 0.85 (0.88) & 0.91 (0.92) & 0.84 (0.88) &0.92 (0.93) & 0.85 (0.88) & 0.91 (0.92) & 0.84 (0.88)& 109 & 99 \\
Layer 24 & 0.91 (0.94) & 0.83 (0.89) & 0.92 (0.92)& 0.84 (0.87) &0.91 (0.94) & 0.83 (0.89) & 0.92 (0.92)& 0.84 (0.87) & 125 & 118 \\
Layer 25 & 0.89 (0.92) & 0.80 (0.86) & 0.89 (0.92) & 0.79 (0.86) & 0.89 (0.92) & 0.80 (0.86) & 0.89 (0.92) & 0.79 (0.86)& 110 &  100\\
Layer 26 & 0.92 (0.93) & 0.82 (0.87)& 0.91 (0.92) & 0.82 (0.85) & 0.92 (0.93) & 0.82 (0.87)& 0.91 (0.92) & 0.82 (0.85)& 110 &  92\\
Layer 27 & 0.88 (0.92) & 0.78 (0.85) &0.90 (0.91) & 0.80 (0.84)&0.88 (0.92) & 0.78 (0.85) &0.90 (0.91) & 0.80 (0.84) & 101 &  84\\
Layer 28 & 0.93 (0.94) & 0.84 (0.90)& 0.93 (0.93)& 0.87 (0.88) & 0.93 (0.94) & 0.84 (0.90)& 0.93 (0.93)& 0.87 (0.88)&112 & 92 \\
Layer 29 & 0.91 (0.90) &0.83 (0.82) & 0.91 (0.90) & 0.82 (0.83) &0.91 (0.90) &0.83 (0.82) & 0.91 (0.90) & 0.82 (0.83)& 95 & 68 \\
Layer 30 & 0.88  (0.89) & 0.77 (0.82)  & 0.89 (0.88) & 0.79 (0.81) &0.88  (0.89) & 0.77 (0.82)  & 0.89 (0.88) & 0.79 (0.81)& 91 & 63\\
Layer 31 & 0.83 (0.82) & 0.69 (0.73) & 0.82 (0.82) & 0.70 (0.72) &0.83 (0.82) & 0.69 (0.73) & 0.82 (0.82) & 0.70 (0.72)& 84& 37 \\
Layer 32 &0.84 (0.89) &0.73 (0.81) & 0.86 (0.87) &0.76 (0.80) &0.84 (0.89) &0.73 (0.81) & 0.86 (0.87) &0.76 (0.80)&125 & 64 \\

\hline
\end{tabular}
}
\caption{Accuracy and macro-avg F1 scores of the test set (training set) for CAV and CAR methods, before and after neuron selection by the chi-squared test for K=50}
\label{tab:large_table1}
\end{table*}

\clearpage
\begin{table*}[ht]
\centering
\small
\setlength{\tabcolsep}{3pt}
\renewcommand{\arraystretch}{1.2}
\begin{tabular}{|c|c|c|c|c|c|c|c|c|c|c|}

\hline
Layer & CAV ACC & CAV F1 & CAR ACC & CAR F1 & CAV-Sel ACC & CAV-Sel F1 & CAR-Sel ACC & CAR-Sel F1 & \# N & \# Sel N \\
\hline
Layer1  & 0.77 (0.82) & 0.71 (0.77) & 0.73 (0.74) & 0.65 (0.71) & 0.77 (0.82) & 0.71 (0.77) & 0.73 (0.74) & 0.65 (0.71) & 223 & 192 \\
Layer2  & 0.74 (0.77) & 0.68 (0.75) & 0.75 (0.85) & 0.68 (0.86) & 0.74 (0.77) & 0.68 (0.75) & 0.75 (0.85) & 0.68 (0.86) & 201 & 172 \\
Layer3  & 0.70 (0.78) & 0.64 (0.75) & 0.69 (0.73) & 0.61 (0.73) & 0.70 (0.78) & 0.64 (0.75) & 0.69 (0.73) & 0.61 (0.73) & 202 & 160 \\
Layer4  & 0.79 (0.81) & 0.73 (0.78) & 0.75 (0.77) & 0.71 (0.75) & 0.79 (0.81) & 0.73 (0.78) & 0.75 (0.77) & 0.71 (0.75) & 198 & 162 \\
Layer5  & 0.73 (0.82) & 0.65 (0.80) & 0.72 (0.77) & 0.68 (0.75) & 0.79 (0.81) & 0.73 (0.78) & 0.72 (0.77) & 0.68 (0.75) & 217 & 186 \\
Layer6  & 0.75 (0.83) & 0.68 (0.80) & 0.73 (0.78) & 0.67 (0.74) & 0.75 (0.83) & 0.68 (0.80) & 0.73 (0.78) & 0.67 (0.74) & 211 & 172 \\
Layer7  & 0.76 (0.82) & 0.69 (0.80) & 0.75 (0.78) & 0.70 (0.77) & 0.76 (0.82) & 0.69 (0.80) & 0.75 (0.78) & 0.70 (0.77) & 226 & 179 \\
Layer8  & 0.71 (0.79) & 0.61 (0.75) & 0.71 (0.76) & 0.64 (0.73) & 0.71 (0.79) & 0.61 (0.75) & 0.71 (0.76) & 0.64 (0.73) & 209 & 153 \\
Layer9  & 0.71 (0.79) & 0.60 (0.72) & 0.72 (0.77) & 0.60 (0.70) & 0.71 (0.79) & 0.60 (0.72) & 0.72 (0.77) & 0.60 (0.70) & 174 & 112 \\
Layer10 & 0.69 (0.76) & 0.61 (0.71) & 0.68 (0.73) & 0.58 (0.68) & 0.69 (0.76) & 0.61 (0.71) & 0.68 (0.73) & 0.58 (0.68) & 183 & 120 \\
Layer11 & 0.70 (0.78) & 0.60 (0.74) & 0.72 (0.77) & 0.64 (0.72) & 0.70 (0.78) & 0.60 (0.74) & 0.72 (0.77) & 0.64 (0.72) & 183 & 110 \\
Layer12 & 0.78 (0.81) & 0.70 (0.75) & 0.76 (0.81) & 0.68 (0.77) & 0.78 (0.81) & 0.70 (0.75) & 0.76 (0.81) & 0.68 (0.77) & 196 & 118 \\
Layer13 & 0.81 (0.87) & 0.70 (0.80) & 0.81 (0.84) & 0.71 (0.79) & 0.81 (0.87) &  0.70 (0.80) & 0.81 (0.84) & 0.71 (0.79) & 180 & 135 \\
Layer14 & 0.84 (0.88) & 0.73 (0.83) & 0.85 (0.85) & 0.75 (0.80) & 0.84 (0.88) & 0.73 (0.83) & 0.85 (0.85) & 0.75 (0.80) & 196 & 164 \\
Layer15 & 0.82 (0.91) & 0.71 (0.86) & 0.86 (0.89) & 0.79 (0.83) & 0.82 (0.91) & 0.71 (0.86) & 0.86 (0.89) & 0.79 (0.83) & 202 & 182 \\
Layer16 & 0.89 (0.92) & 0.80 (0.87) & 0.89 (0.91) & 0.80 (0.86) & 0.89 (0.92) & 0.80 (0.87) & 0.89 (0.91) & 0.80 (0.86) & 203 & 184 \\
Layer17 & 0.88 (0.91) & 0.77 (0.86) & 0.87 (0.90) & 0.77 (0.84) & 0.88 (0.91) & 0.77 (0.86) & 0.87 (0.90) & 0.77 (0.84) & 208 & 193 \\
Layer18 & 0.90 (0.93) & 0.80 (0.89) & 0.89 (0.92) & 0.80 (0.88) & 0.90 (0.93) & 0.80 (0.89) & 0.89 (0.92) & 0.80 (0.88) & 198 & 186 \\
Layer19 & 0.89 (0.94) & 0.77 (0.90) & 0.90 (0.92) & 0.78 (0.88)  & 0.89 (0.94) & 0.77 (0.90) & 0.90 (0.92) & 0.78 (0.88) & 211 & 201 \\
Layer20 & 0.91 (0.94) & 0.83 (0.89) & 0.90 (0.92) & 0.81 (0.87) & 0.91 (0.94) & 0.83 (0.89) & 0.90 (0.92) & 0.81 (0.87) & 209 & 193 \\
Layer21 & 0.93 (0.95) & 0.87 (0.92) & 0.92 (0.93) & 0.86 (0.90) & 0.93 (0.95) & 0.87 (0.92) & 0.92 (0.93) & 0.86 (0.90) & 242 & 227 \\
Layer22 & 0.94 (0.96) & 0.87 (0.93) & 0.93 (0.95) & 0.84 (0.92) & 0.94 (0.96) & 0.87 (0.93) & 0.93 (0.95) & 0.84 (0.92) & 223 & 215 \\
Layer23 & 0.94 (0.95) & 0.89 (0.92) & 0.92 (0.94) & 0.86 (0.90) & 0.94 (0.95) & 0.89 (0.92) & 0.92 (0.94) & 0.86 (0.90) & 235 & 229 \\
Layer24 & 0.93 (0.95) & 0.88 (0.91) & 0.94 (0.94) & 0.88 (0.91) & 0.93 (0.95) & 0.88 (0.91) & 0.94 (0.94) & 0.88 (0.91) & 246 & 238 \\
Layer25 & 0.93 (0.95) & 0.85 (0.92) & 0.92 (0.94) & 0.84 (0.91) & 0.93 (0.95) & 0.85 (0.92) & 0.92 (0.94) & 0.84 (0.91) & 244 & 232\\
Layer26 & 0.92 (0.96) & 0.83 (0.92) & 0.93 (0.94) & 0.86 (0.90) & 0.92 (0.96) & 0.83 (0.92) & 0.93 (0.94) & 0.86 (0.90) & 219 & 211 \\
Layer27 & 0.91 (0.95) & 0.81 (0.91) & 0.91 (0.94) & 0.83 (0.89) & 0.91 (0.95) & 0.81 (0.91) & 0.91 (0.94) & 0.83 (0.89) & 218 & 201 \\
Layer28 & 0.94 (0.96) & 0.88 (0.93) & 0.95 (0.94) & 0.90 (0.91) & 0.94 (0.96) & 0.88 (0.93) & 0.95 (0.94) & 0.90 (0.91) & 232 & 210 \\
Layer29 & 0.93 (0.94) & 0.87 (0.90) & 0.93 (0.93) & 0.85 (0.89) & 0.93 (0.94) & 0.87 (0.90) & 0.93 (0.93) & 0.85 (0.89) & 192 & 167 \\
Layer30 & 0.91 (0.94) & 0.80 (0.90) & 0.91 (0.94) & 0.78 (0.90) & 0.91 (0.94) & 0.80 (0.90) & 0.91 (0.94) & 0.78 (0.90) & 190 & 174 \\
Layer31 & 0.91 (0.92) & 0.83 (0.86) & 0.91 (0.91) & 0.82 (0.85) & 0.91 (0.92)  & 0.83 (0.86) & 0.91 (0.91) & 0.82 (0.85) & 150 & 125 \\
Layer32 & 0.90 (0.94) & 0.82 (0.91) & 0.91 (0.92) & 0.84 (0.89) & 0.90 (0.94) & 0.82 (0.91) & 0.91 (0.92) & 0.84 (0.89) & 202 & 177 \\

\hline
\end{tabular}

\caption{Accuracy and macro-avg F1 scores of the test set (training set) for CAV and CAR methods, before and after neuron selection by the test of independence for K=100. Results of goodness-of-fit is shown in the paper.}
\label{tab:large_table1}
\end{table*}
\end{document}